\documentclass[sigconf]{acmart}
\AtBeginDocument{%
  }

\setcopyright{acmlicensed}
\copyrightyear{2027}
\acmYear{2027}
\acmDOI{XXXXXXX.XXXXXXX} 
\acmConference[Conference'26]{Conference 2026}{date}{place}
\acmISBN{978-1-4503-XXXX-X/2018/06}

\usepackage{amsmath}
\usepackage{booktabs}
\usepackage{graphicx}
\usepackage{subcaption}
\usepackage{listings}
\usepackage{xcolor}
\usepackage{paralist}
\usepackage{enumitem}
\usepackage{amsmath}
\usepackage{xcolor}
\usepackage{fontawesome5}
\usepackage{tabularx}
\usepackage{xspace}
\usepackage{tcolorbox}
\tcbuselibrary{skins,breakable}
\usepackage{mdframed}

\newcommand{\system}{\textsc{MapDB}\xspace}

\definecolor{sqlblue}{RGB}{26,72,125}
\definecolor{sqlgray}{RGB}{90,90,90}
\definecolor{sqlred}{RGB}{130,40,40}
\definecolor{mapblue}{RGB}{24,82,150}

\definecolor{alertbg}{RGB}{245, 245, 245}
\definecolor{alertborder}{RGB}{166, 198, 252}

\lstdefinestyle{sqlstyle}{
    language=SQL,
    basicstyle=\ttfamily\scriptsize,
    keywordstyle=\bfseries\color{sqlblue},
    commentstyle=\itshape\color{sqlgray},
    stringstyle=\color{sqlred},
    breaklines=true,
    frame=single,
    columns=fullflexible,
    showstringspaces=false,
    xleftmargin=1pt,
    xrightmargin=1pt
}

\newtcolorbox{examplebox}[1]{
  colback=alertbg, colframe=alertborder,
  fonttitle=\small\bfseries\color{black},
  title={#1},
  left=4pt, right=4pt, top=3pt, bottom=3pt,
  arc=2pt, boxrule=0.8pt,
  before skip=6pt, after skip=6pt
}

\newcommand{\ie}{\textit{i.e.},\ }

\begin{document}

\title{Queryable Self-Organizing Maps: A Database Abstraction for Topology-Driven Data Exploration
}



\author{Denis Mayr Lima Martins}
\authornote{Both authors contributed equally to this research.}
\affiliation{%
  \institution{Department of Computing and Mathematics\\University of Sao Paulo}
  \city{Ribeirão Preto}
  \state{São Paulo}
  \country{Brazil}
}\email{martins.denis@usp.br}
\orcid{0000-0002-8262-2369}

\author{Gottfried Vossen}
\affiliation{%
  \institution{Department of Information Systems\\University of M{\"u}nster}
  \city{M{\"u}nster}
  \country{Germany}}
\email{vossen@uni-muenster.de}

\renewcommand{\shortauthors}{Martins et al.}

\begin{abstract}
  Self-Organizing Maps (SOMs) have long been used as exploratory tools for high-dimensional data: they organize objects into a two-dimensional topology that reveals clusters, gradients, sparse regions, dense regions, and boundaries. Yet, in modern data systems, SOMs are typically trained and visualized outside the DBMS, disconnected from the relational data they summarize. We introduce the abstraction of a \emph{queryable data map}: a learned topological artifact consisting of representatives, neighborhood relations, object assignments, and derived summaries. We instantiate this idea with \system, a lightweight prototype that makes SOM artifacts queryable so users can explore data topology without leaving the database. Experimental study shows that SOM training is feasible at moderate analytical scale, that map queries are interactive after materialization, and that SOM regions provide meaningful targets for exploratory SQL.
\end{abstract}

\received{20 February 2007}
\received[revised]{12 March 2009}
\received[accepted]{5 June 2009}

\maketitle

\section{Introduction}

Before analysts can write precise SQL predicates, they often need to understand the shape of the data. This is difficult in high-dimensional relational datasets, where clusters, gradients, outliers, and boundaries may involve many interacting attributes. We fuse the concept of Self-Organizing Maps (SOMs)~\cite{kohonen1990self,martins2023som,carvalho2022som} with relational databases in MapDB, making the former intensional data that can be queried just like other databse objects.

\begin{examplebox}{Motivating Example}\label{ex:motivation}
\small
Consider an analyst investigating a customer table with dozens of behavioral attributes, or a customer-level analytical view derived from \texttt{customer}, \texttt{orders}, \texttt{lineitem}, and \texttt{supplier}. The analyst may not yet know which predicates are useful, which combinations of attributes define recurring profiles, where unusual objects lie, or whether a transition between two populations is sharp or gradual. Conventional SQL assumes that the analyst already knows what to filter, group, or compare. This leads to an important \emph{gap between relational querying and exploratory understanding}.
\end{examplebox}

SOMs are a compelling example of a structure that addresses this gap. A SOM organizes high-dimensional data over a low-dimensional lattice while encouraging nearby map units to represent similar input profiles~\cite{kohonen1990self}, 
As a result, SOMs have been widely used as exploratory data maps~\cite{vesanto2000clustering,ultsch2005clustering}, where:

\begin{itemize}
    \item \textbf{hit maps} reveal dense and sparse areas;
    \item \textbf{Unified-Distance- or U-Matrices} reveal local boundaries;
    \item \textbf{component planes} reveal feature gradients;
    \item \textbf{prototype vectors} summarize local profiles; and
    \item \textbf{assignments} connect objects to topological regions.
\end{itemize}

Despite this exploratory role, SOMs are usually produced outside the database system as transient Python, R, or visualization artifacts. The source data remains queryable, but the map that explains the data is not. This separation is harmful: map regions cannot be naturally joined with source tuples; map versions are not governed by the DBMS; and exploratory recommendations cannot directly exploit the learned topology. If a SOM helps an analyst decide where to look, then the DBMS should know what the SOM contains.

This paper proposes that SOMs should be treated as \emph{intensional data}: persistent, learned topological artifacts stored and queried inside the DBMS. The key observation is that a SOM is almost embarrassingly relational. It consists of neurons, prototype vectors, topology edges, object-to-neuron assignments, and derived summaries such as hit counts, U-Matrix values, and component planes. Each of these objects can be represented as a relation. 



We present \system, a lightweight prototype that stores and queries SOMs inside a relational database. \system does not aim to outperform specialized SOM libraries. Instead, it demonstrates a database abstraction: exploratory maps can be stored, queried, joined, versioned, and reused like other database artifacts. This enables SQL queries over specific regions of the map, and finding source tuples that are mapped near a selected neuron.



The paper makes four contributions:

\begin{enumerate}
    \item We introduce \emph{learned maps as intensional data}, arguing that persistent topology can provide a queryable organization over relational tuples.
    \item We present \system, a relational representation and SQL-resident execution substrate for SOM neurons, prototypes, topology, assignments, and summaries.
    \item We introduce a compact family of map-aware SQL extensions for map creation, density analysis, U-Matrix inspection, component-plane queries, neighborhood traversal, region drill-down, boundary discovery, and selective summary materialization.
    \item We provide a comprehensive evaluation covering scalability,
    flat-clustering baselines, materialization trade-offs, map-size sensitivity, TPC-H scaling, and single-table and multi-table case studies.
\end{enumerate}


\section{From Visual SOMs to Queryable Data Maps}


SOMs are typically consumed as visualizations. A user inspects a two-dimensional map and manually interprets dense regions, boundaries, and feature gradients. However, a visualization alone is not enough for database exploration. A useful exploratory map should support selection, filtering, joining, aggregation, versioning, comparison, and provenance. These are database responsibilities.

We therefore define a \emph{queryable data map} as a learned topological artifact:
\begin{equation}
    \mathcal{M} = (V, E, W, A, S),
    \label{eq:data-map}
\end{equation}

\noindent
where $V$ is a set of map units, $E \subseteq V \times V$ is a neighborhood relation, $W$ stores representative parameters, $A$ assigns database objects to map units, and $S$ stores derived summaries over the map.

For a SOM, $V$ is the set of neurons, $E$ is the lattice topology, $W$ is the set of prototype vectors, $A$ is the best-matching-unit assignment relation, and $S$ includes U-Matrix values, hit counts, component planes, and local error summaries. The important point is that $\mathcal{M}$ is both visual and relational. It supports visual exploration because it can be rendered as a map. It supports database exploration because each component is queryable.

This abstraction is related to the broader view that learned models can be treated as database artifacts rather than opaque external objects~\cite{10.1145/3735654.3735946}. However, SOMs add an exploratory dimension. They are not only functions to evaluate; they are maps that organize the data space. This makes them particularly suitable for data exploration, workload understanding, and topology-aware recommendation.

\section{\system Design: SOMs as Relational Objects}
\label{sec:design}

\begin{figure}[t]
\centering
\includegraphics[width=\linewidth]{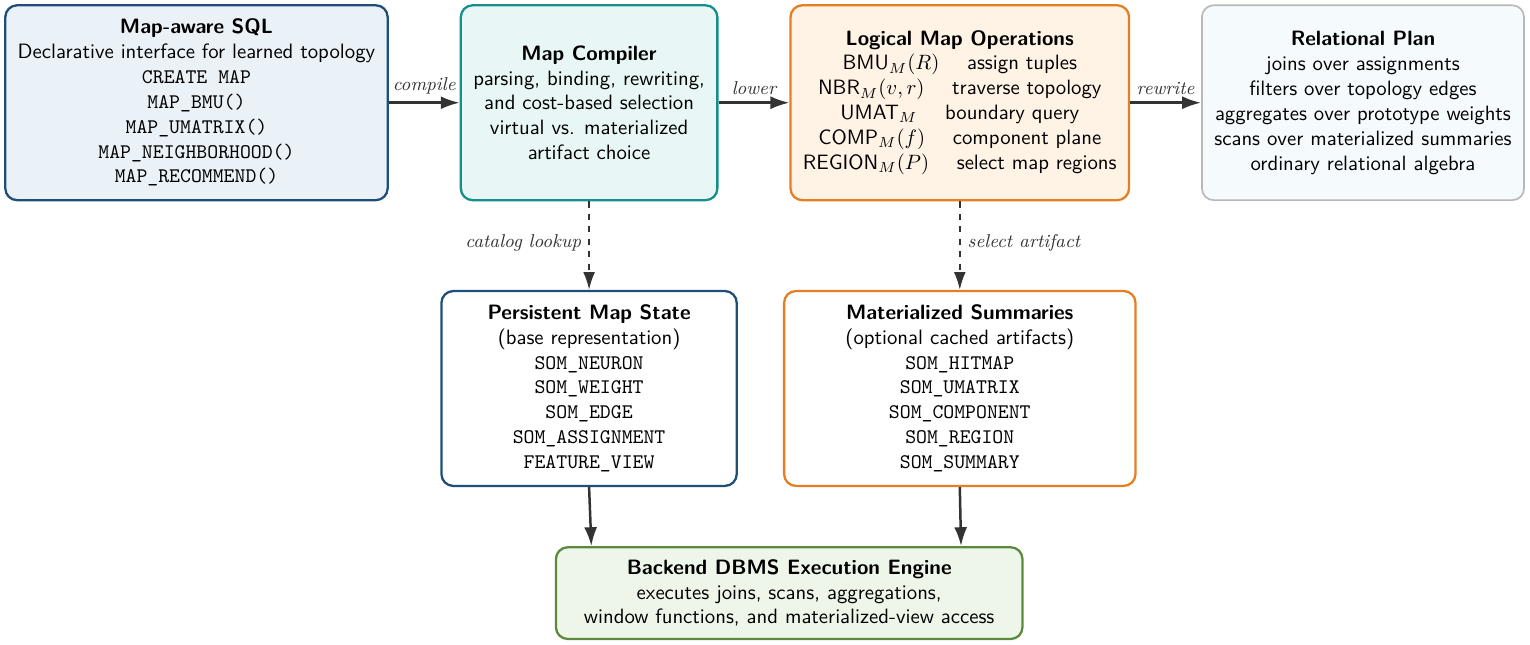}
\caption{
    \system architecture. Map-aware SQL operations compile to logical map
    operations and then to relational plans over persistent map state or
    materialized summaries. The backend DBMS executes the resulting plan.
}
\label{fig:architecture}
\end{figure}

Figure~\ref{fig:architecture} summarizes the architecture. In essence, the design separates:
\begin{inparaenum}
    \item map-aware SQL syntax;
    \item logical map operations;
    \item persistent map state; and
    \item physical relational strategies. 
\end{inparaenum} 
The schema is intentionally simple and includes five core relations (see Table~\ref{tab:schema}).

\begin{table}[t]
\centering
\caption{\system core relations.}
\label{tab:schema}
\small
\begin{tabular}{ll}
\toprule
\textbf{Relation} & \textbf{Meaning} \\
\midrule
\texttt{MAP\_NEURON} & map units and positions \\
\texttt{MAP\_WEIGHT} & prototype values \\
\texttt{MAP\_EDGE} & topology relation \\
\texttt{MAP\_ASSIGNMENT} & object-to-unit mapping \\
\texttt{MAP\_SUMMARY} & derived summaries \\
\bottomrule
\end{tabular}
\end{table}

Conceptually:

\begin{lstlisting}[style=sqlstyle, caption={Core Relations.},label={lst:core}]
MAP_NEURON(map_id, neuron_id, row_pos, col_pos)
MAP_WEIGHT(map_id, epoch, neuron_id, feature_id, value)
MAP_EDGE(map_id, src_neuron, dst_neuron, grid_distance)
MAP_ASSIGNMENT(map_id, epoch,
object_id, neuron_id, distance)
MAP_SUMMARY(map_id, epoch, neuron_id, summary_name, value)
\end{lstlisting}

A map catalog additionally stores:

\begin{lstlisting}[style=sqlstyle]
MAP_CATALOG(map_id, map_name, source_relation, object_id_column, rows, cols, current_epoch, status)

MAP_FEATURE(map_id, feature_id, feature_name, transform_spec)
\end{lstlisting}

The input data are represented as a long feature relation:

\begin{lstlisting}[style=sqlstyle]
MAP_INPUT_LONG(object_id, feature_id, value)
\end{lstlisting}

This representation is flexible and uniform: the same SQL patterns support different datasets and feature spaces. Dense numerical datasets can also be stored in wide format or vector columns; we use the long format because it exposes the relational structure most clearly.

\paragraph{\bf Training}

For input object object $x_i$, the Best Matching Unit (BMU) is:
\begin{equation}
    b_i
    =
    \arg\min_j
    \lVert
    x_i-w_j
    \rVert_2^2.
    \label{eq:bmu}
\end{equation}

\noindent 
\system computes assignments using relational joins, aggregation, and window ranking. A simplified form is:

\begin{lstlisting}[style=sqlstyle, caption={BMU assignment in SQL.},label={lst:bmu}]
SELECT object_id, neuron_id, dist
FROM (
  SELECT
    x.object_id, w.neuron_id,
    SUM(POWER(x.value-w.value,2)) AS dist,
    ROW_NUMBER() OVER (
      PARTITION BY x.object_id
      ORDER BY
        SUM(POWER(x.value-w.value,2))
    ) AS rn
  FROM MAP_INPUT_LONG x
  JOIN MAP_WEIGHT w ON x.feature_id = w.feature_id
  WHERE w.map_id = :map AND w.epoch = :epoch
  GROUP BY
    x.object_id,
    w.neuron_id
) q
WHERE rn = 1;
\end{lstlisting}

\system uses batch SOM training because it is naturally relational. Given BMU assignments, each prototype is recomputed as a neighborhood-weighted average:

\begin{equation}
    w_j^{t+1}
    =
    \frac{
        \sum_i
        h_{b_i,j}^{t}
        x_i
    }{
        \sum_i
        h_{b_i,j}^{t}
    },
    \label{eq:batch-update}
\end{equation}

with:

\begin{equation}
    h_{b_i,j}^{t}
    =
    \exp
    \left(
    -
    \frac{
        d_{\mathrm{grid}}(b_i,j)^2
    }{
        2\sigma_t^2
    }
    \right).
    \label{eq:neighborhood}
\end{equation}

The update joins assignments, topology edges, and input features, then aggregates by target neuron and feature. We highlight that training inside SQL is not the end goal. It is the mechanism that keeps the learned topology inside the data system, where it can be queried, versioned, and joined with source data.

\smallskip

\paragraph{\bf \system SQL Extensions}
\label{sec:sql-extensions}
Once the SOM is represented relationally, model analysis becomes SQL. However, the low-level relations are useful for transparency and execution, but they are not an appropriate user abstraction. Therefore, \system exposes a compact family of map-aware SQL extensions.

\paragraph{Map lifecycle}

A map can be created from a relation or analytical view:

\begin{lstlisting}[style=sqlstyle]
CREATE MAP census_map USING SOM ON census_features
OBJECT ID object_id
FEATURES (age, education_num, hours_per_week, capital_gain_log1p)
WITH (rows = 10, cols = 10, epochs = 20);
\end{lstlisting}

Additional lifecycle statements include:

\begin{lstlisting}[style=sqlstyle]
TRAIN MAP census_map
FOR 20 EPOCHS;
\end{lstlisting}

\begin{lstlisting}[style=sqlstyle]
REFRESH MAP census_map;
\end{lstlisting}

\begin{lstlisting}[style=sqlstyle]
DROP MAP census_map;
\end{lstlisting}

\paragraph{Density Queries}

The logical operation:

\begin{lstlisting}[style=sqlstyle]
SELECT * FROM mapdb.map_hits('census_map');
\end{lstlisting}

returns neuron occupancy.

Dense neurons reveal common data profiles and are queried as:

\begin{lstlisting}[style=sqlstyle,caption={Dense-region discovery.},label={lst:hitmap}]
SELECT * FROM mapdb.map_dense_regions('census_map', 100);
\end{lstlisting}

\paragraph{Boundary Queries}

The U-Matrix~\cite{ultsch2005clustering} is a common SOM visualization that identifies neurons whose prototypes differ strongly from neighboring prototypes. 
For neuron $j$, the U-Matrix value is:

\begin{equation}
    U(j)
    =
    \frac{1}{
        |\mathcal{N}(j)|
    }
    \sum_{
        k\in\mathcal{N}(j)
    }
    \lVert
    w_j-w_k
    \rVert_2.
    \label{eq:umatrix}
\end{equation}

Users can query it through:

\begin{lstlisting}[style=sqlstyle]
SELECT * FROM mapdb.map_umatrix('census_map');
\end{lstlisting}

or select strong boundaries:

\begin{lstlisting}[style=sqlstyle,caption={Boundary-neuron query.},label={lst:boundary}]
SELECT * FROM mapdb.map_boundaries('census_map', 0.8);
\end{lstlisting}

\paragraph{Component Queries}

Component planes show how individual features vary across the map (\ie, a feature-specific projection of the prototypes):

\begin{lstlisting}[style=sqlstyle,caption={Component-plane query.},label={lst:component}]
SELECT * FROM mapdb.map_component('census_map', 'education_num');
\end{lstlisting}

This query returns: $(row,\;column,\;value)$. Thus component-plane figures are views over persistent model state rather than external plotting artifacts.

\paragraph{Topology Traversal}

The most important database interaction is region-to-data drill-down. A user can select a neuron or connected region and retrieve the corresponding source records. Neighborhood traversal is exposed through:

\begin{lstlisting}[style=sqlstyle]
SELECT * FROM mapdb.map_neighborhood'census_map', 42, 2);
\end{lstlisting}

Immediate adjacency is:

\begin{lstlisting}[style=sqlstyle]
SELECT * FROM mapdb.map_adjacent('census_map', 42);
\end{lstlisting}

The result can be joined with the source relation:

\begin{lstlisting}[style=sqlstyle,caption={Retrieving source tuples by map neighborhood.},label={lst:region-drilldown}]
SELECT a.* FROM census a JOIN mapdb.map_region_objects('census_map', 42, 2) r ON a.object_id = r.object_id;
\end{lstlisting}

This operation retrieve all records mapped near a selected region of the learned topology. It turns visual navigation into database navigation.

\section{Experimental Analysis}

Our evaluation mainly concerns the scalability of SOM processing in terms of data size, feature dimensionality, and map size. In addition, we evaluate the speedup gains by materializing map artifacts. We also included experiments to check whether the abstraction remain useful for both single-table and multi-table relational settings.

\paragraph{Prototype}

MapDB is implemented in DuckDB with Python orchestration. DuckDB stores the relational SOM artifacts and executes BMU assignment, batch updates, map queries, materialization, and case-study drill-down. Python manages data loading, feature preprocessing, repeated runs, baseline models, plotting, and table generation.

\paragraph{Datasets} We employ three dataset in our experiments:
\begin{itemize}
    \item \textbf{Synthetic}: We generate controlled Gaussian-mixture data and vary the number of objects $N$, dimensions $D$, and map units $K$.
    \item \textbf{Adult Census}: Numeric features are imputed when necessary, skewed attributes are log-transformed, and numeric values are standardized. Categorical attributes are one-hot encoded; missing categorical values are represented explicitly. To prevent high-cardinality groups from dominating Euclidean distance, dummy variables belonging to an original categorical attribute with $K_g$ categories are weighted by $1/\sqrt{K_g}$. Income is excluded from training and used only for interpretation.
    \item \textbf{TPC-H}: We construct customer-level feature views through joins and aggregation over \texttt{customer}, \texttt{orders}, \texttt{lineitem}, \texttt{supplier}, and \texttt{part}. Features include order counts, total spend, average order value, quantity, discount, line price, shipping delay, supplier diversity, and part diversity. Skewed aggregates are log-transformed and standardized. Market segment is retained as an external interpretation label.
\end{itemize}




\paragraph{Baselines and Metrics}
We compare MapDB against our own Python implementation of Batch SOM, KMeans, and a mini-batch version of KMeans. For fair comparison, clustering methods use the same number of representatives $K = \text{map rows} \times \text{map columns}$. For evaluating the quality of each method, we report quantization error and silhouette score.

\subsection{Results}

\begin{figure}[t]
    \centering  \includegraphics[width=\linewidth]
        {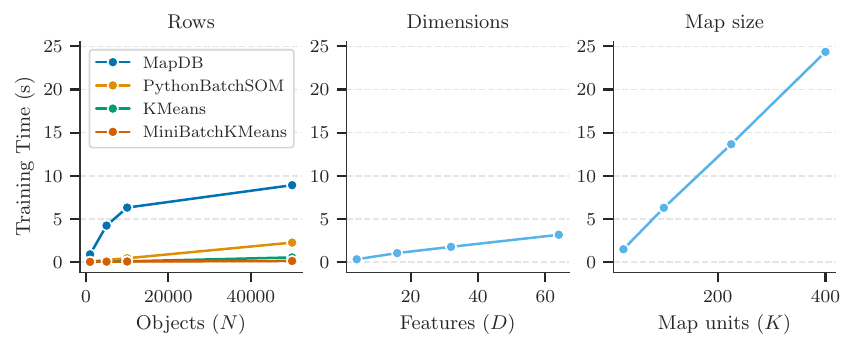} 
    \caption{
        Synthetic scalability as the number of objects, dimensions, and
        map units increase.
    }
    \label{fig:scalability}
\end{figure}

\paragraph{Synthetic Scalability}
We first characterize generic runtime behavior under controlled variation of $N$, $D$, and map size (in terms of the number of map units $K$). 
Figure~\ref{fig:scalability} shows that total training time increases with all three factors, as expected from repeated object-to-prototype comparisons. The dominant component is BMU assignment. At the largest tested configuration, MapDB requires approximately 20 seconds to training a map. 

This result demonstrates that relational SOM processing is feasible for moderate analytical workloads, but larger configurations and larger database sizes motivate vectorized or approximate BMU search.

\paragraph{SOM versus Flat Clustering}

\begin{figure}[t]
    \centering
    \includegraphics[width=\linewidth]
        {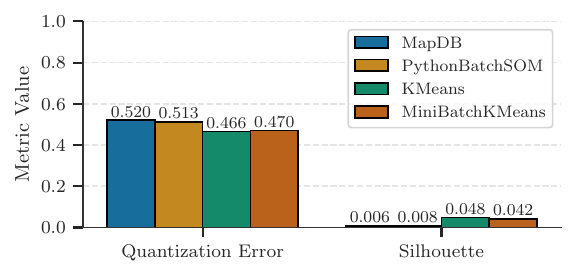}
    \caption{
        SOM and flat clustering baselines. The comparison includes
        representation quality and region-quality metrics.
    }
    \label{fig:baseline}
\end{figure}

A central question in our work is whether SOM provides value beyond simply assigning tuples to clusters. 
Figure~\ref{fig:baseline} compares MapDB, Python Batch SOM, KMeans, and MiniBatchKMeans. 

MapDB achieves competivitive results (while taking more runtime, as described before). Flat clustering remains competitive on the Silhouette metric, but lacks explicit adjacency and therefore cannot directly support neighborhood expansion, U-Matrix boundaries, or topological traversal. 

\paragraph{Materialized versus Recomputed Queries}



\begin{figure}[t]
    \centering
    \includegraphics[width=\linewidth]
        {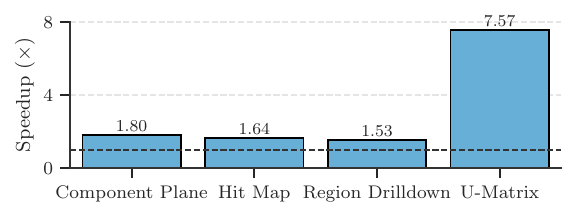}
    \caption{
    Median speedup gains from artifact materialization.
    }
    \label{fig:materialization}
\end{figure}

We benchmark hit maps, U-Matrix values, component planes, and region drill-down. Before timing, MapDB checks that both modes return equivalent results. 
Figure~\ref{fig:materialization} the speedup obtained by querying materialized SOM artifacts instead of recomputing the same artifacts
from lower-level map state. Materialization improves all tested query types, with U-Matrix queries showing the largest benefit, with a median speedup of $7.57$. This is due to the intrisinc complexity of producing the U-Matrix: recomputation requires joins over topology edges and prototype weights, followed by distance aggregation over neighboring neurons. In contrast, materialized U-Matrix access reduces the operation to a scan over a compact summary relation. 

These results indicate the need for a selective materialization strategy in which the database system should persist expensive or frequently accessed map-derived
summaries while leaving cheaper artifacts virtual when appropriate.




\subsection{Case Studies}

The systems experiments above test feasibility and topology. We now ask whether queryable maps expose useful exploratory structures in real relational settings.

\paragraph{Adult Dataset: Single-Table Exploration}

Adult represents a single-table mixed-type setting. The SOM is trained without the income label. After training, map regions are joined back to source tuples and external labels for interpretation.

\begin{figure}[t]
    \centering
    \includegraphics[width=\linewidth]{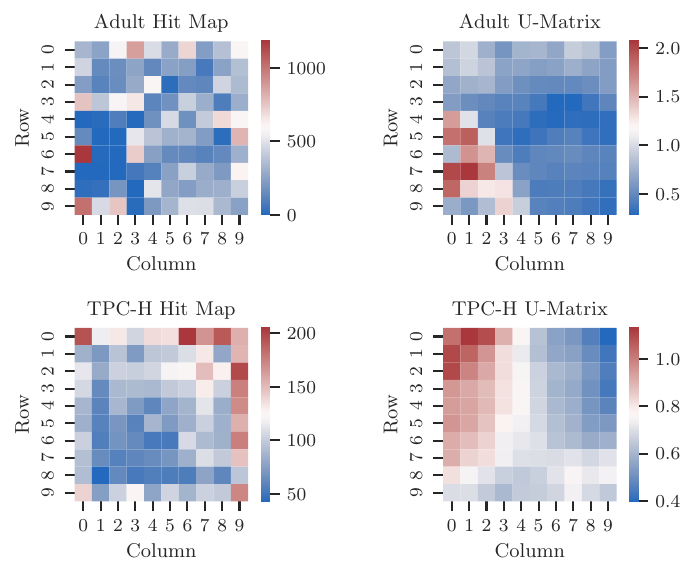}
    \caption{
        Case-study artifacts. Adult exposes non-uniform density and strong topological boundaries. TPC-H exposes structured occupancy over a multi-table analytical view and a coherent feature gradient.
    }
    \label{fig:cases}
\end{figure}

The Adult hit map in Figure~\ref{fig:cases} shows strongly non-uniform occupancy, in which recurring demographic and employment profiles concentrate in localized areas, while other regions remain sparsely populated. The U-Matrix reveals pronounced high-distance zones on one side of the topology and smoother transitions elsewhere. 

The combination of density and boundary strength is especially useful. A populated region near a strong U-Matrix boundary is a natural target for comparison because it contains many records while lying close to a sharp profile transition.

\begin{figure}[t]
\centering
\includegraphics[width=.95\linewidth]{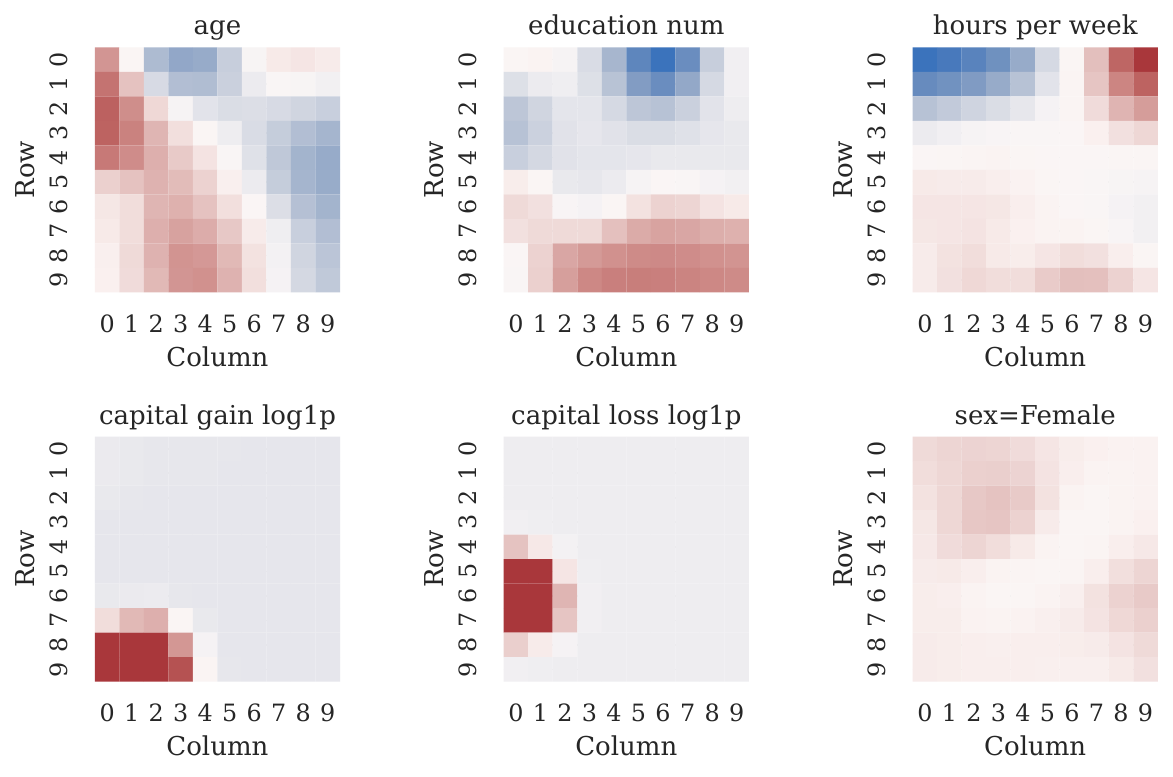}
\includegraphics[width=.95\linewidth]{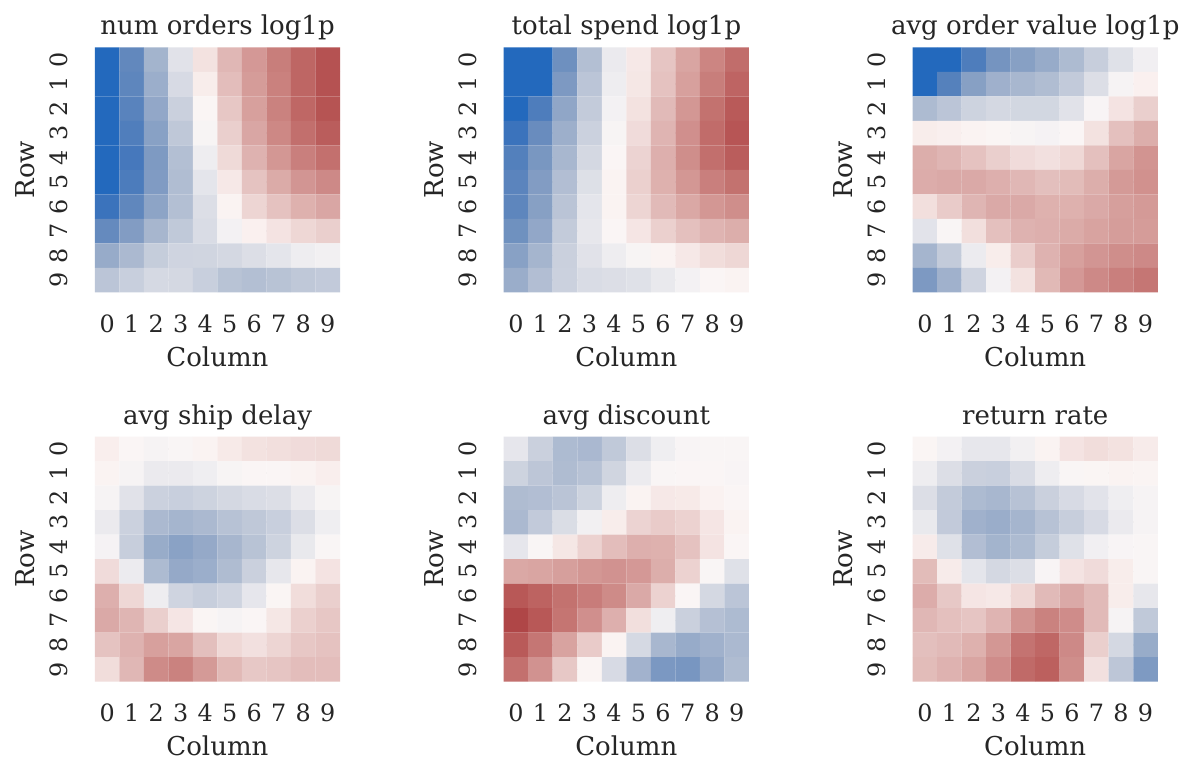}
\caption{
    Selected Adult (top rows) and TPC-H (bottom rows) component planes. Each panel is obtained by querying a feature slice of SOM weights. Smooth gradients and high-value areas expose how features vary over the learned topology.
}
\label{fig:components}
\end{figure}

Component planes in Figure~\ref{fig:components} indicate that the learned topology is meaningful, where neighboring neurons correspond to systematically changing demographic and employment profiles. For instance, age varies mainly from older profiles on the left to younger profiles on the right, while education exhibits a clear top--bottom contrast, with higher values concentrated in the lower part of the map. Capital gain and capital loss are sparse and localized, indicating that these attributes characterize specific subregions rather than the whole map. The \texttt{sex=Female} plane shows a weaker but still spatially localized pattern.


\paragraph{TPC-H: Multi-Table SQL Exploration}

TPC-H tests whether the same abstraction applies when mapped objects are not stored in a single table but produced by a relational query involving joins and aggregation. We construct customer-level feature vectors from:
\begin{equation*}
\texttt{customer}
\Join
\texttt{orders}
\Join
\texttt{lineitem}
\Join
\texttt{supplier}
\Join
\texttt{part}.
\end{equation*}

The resulting hit map in Figure~\ref{fig:cases} again shows strongly non-uniform occupancy. Dense map regions correspond to recurring customer purchasing and logistics profiles, while sparse regions represent less common combinations. In contrast, component planes in Figure~\ref{fig:components}~(bottom) indicate that the map captures coherent gradients over customer-level aggregate behavior. In particular, the number of orders and total spend increase smoothly from left to right, indicating that the SOM organizes customers by overall purchasing intensity. Morevoer, shipping delay, discount, and return rate exhibit more localized structures, suggesting that logistics and pricing behavior distinguish specific subregions rather than the entire topology. Overall, the component planes show that the map preserves interpretable structure in a multi-table SQL-derived feature space.

\paragraph{Region-to-Data Drill-Down}

For each case study, MapDB selects three representative region types, such as, dense region, boundary region, and high-error region. The system then expands each region by a configurable grid radius, joins assignments back to source tuples, and reports the strongest feature contrasts. 

\paragraph{Region drill-down.}

Table~\ref{tab:case-drilldown} reports obtained results. In the
Adult dataset, dense and boundary regions both correspond to large subpopulations with a slight majority of $>50K$ income labels and higher education-related profiles. In TPC-H, the dense region captures a recurring high-activity customer profile with more orders, whereas the boundary and high-error regions identify lower-value customer profiles with lower average order value, total spend, and quantity. These results illustrate the main benefit of \system drill-down, i.e., map regions become queryable database subsets that can be summarized, compared, and interpreted with SQL.

\begin{table}[t]
\centering
\caption{Case-study region drill-down. Regions selected from the SOM are joined
back to the source feature views and summarized by their dominant label and
largest standardized contrasts against the global population.}
\label{tab:case-drilldown}
\footnotesize
\begin{tabular}{llrll}
\toprule
\textbf{Dataset} &
\textbf{Region} &
\textbf{Size} &
\textbf{Dom. label} &
\textbf{Main contrasts vs. global} \\
\midrule
Adult &
Dense &
1332 &
$>50K$ &
$\uparrow$ capital loss; $\uparrow$ education \\

Adult &
Boundary &
1425 &
$>50K$ &
$\uparrow$ capital loss; $\uparrow$ education \\

Adult &
High-error &
1665 &
$>50K$ &
$\uparrow$ capital gain; $\uparrow$ education \\

TPC-H &
Dense &
846 &
BUILDING &
$\uparrow$ orders; $\downarrow$ avg. line price \\

TPC-H &
Boundary &
695 &
FURNITURE &
$\downarrow$  avg. order value, total spend \\

TPC-H &
High-error &
468 &
BUILDING &
$\downarrow$ avg. order value, and quantity \\
\bottomrule
\end{tabular}
\end{table}

\paragraph{Scaling Multi-Table Feature Views}


\begin{table}[t]
\centering
\caption{TPC-H multi-table scaling. MapDB trains SOMs over customer-level feature views derived from joins and aggregations. Here, QE denotes the quantization error.}
\label{tab:tpch-scaling}
\small
\begin{tabular}{rrrrrr}
\toprule
\textbf{SF} &
\textbf{Objects} &
\textbf{Feat. View (s)} &
\textbf{Training (s)} &
\textbf{BMU (s) } &
\textbf{QE} \\
\midrule
0.01 & 1,000  & 0.04 & 1.61 & 0.87 & 1.54 \\
0.10 & 10,000 & 0.22 & 6.12 & 3.62 & 1.59 \\
1.00 & 30,000 & 0.95 & 7.35 & 4.28 & 1.65 \\
\bottomrule
\end{tabular}
\end{table}

 We evaluate scale factors $SF\in\{0.01,0.1,1.0\}$ for TPC-H. In our runs, feature-view construction stayed below one second at scale factor $1.0$, while SOM training  is the dominant cost. 
The main contributor is BMU assignment, which reaches $4.28s$ at $SF=1.0$. Map quality remains stable across scales, where quantization error increases only modestly from $1.54$ to $1.65$. These results
suggest that the relational construction of mapped objects is not the main bottleneck in this setting; future optimization should focus on BMU assignment, prototype updates, and incremental map maintenance.

 

\section{Related Work}
\label{sec:related-work}

\paragraph{Self-Organizing Maps and exploratory analysis.}
A SOM organizes high-dimensional objects over a low-dimensional topology while encouraging neighboring units to represent similar input profiles~\cite{kohonen1990self}. Prior work has used SOM for clustering, visualization, and exploratory analysis~\cite{lotsch2014exploiting}. \system differs from this literature in focus: rather than proposing a new SOM algorithm, we ask what changes when the learned topology becomes persistent, relationally represented, and directly queryable inside a DBMS.

\paragraph{In-database analytics and machine learning.}
Recent work spans in-database model training, declarative machine learning, and systems that expose model state for inspection or querying. Related efforts such as SQL4NN~\cite{10.1145/3735654.3735946} treat learned models as data that can be validated and queried through relational abstractions.
\system is complementary but targets a different class of object: an exploratory learned structure whose internal topology, assignments, and summaries are themselves useful analytical state. 

\paragraph{Query recommendation and data exploration.}
Exploratory database systems help users identify interesting subsets and formulate subsequent queries when the target predicate is not known in advance. For example, PyExplore derives query recommendations from structure discovered in query results~\cite{10.1145/3448016.3452762}. \system shares the goal of
supporting exploration but introduces a persistent intermediate abstraction:
a learned topological map that can be reused across sessions, traversed by
neighborhood, queried for boundaries or density, and joined back to source
objects. Thus, exploration is expressed over a durable learned organization
rather than only over transient clustering or visualization output.

To our knowledge, the combination explored here (persistent relational representation of SOM topology, map-aware SQL operations, region-to-data
drill-down) has received little attention. \system therefore connects three previously separate concerns: exploratory topological learning, database-native
model state, and declarative relational analytics.

\section{Conclusion}

This paper argues that learned exploratory maps, derived from a Self-Organized Map, should be treated as intensional database objects. A SOM organizes high-dimensional data into a two-dimensional lattice that exposes density, boundaries, gradients, and recurring data profiles. In conventional workflows, this organization remains external to the DBMS. \system instead treats SOM artifacts as persistent database state.

More importantly, \system exposes map-aware SQL operations for creating maps, inspecting component planes, traversing neighborhoods, drilling from learned regions back to source tuples, and selectively materializing expensive summaries.

Our evaluation shows how this abstraction operates across controlled scalability workloads, flat-clustering comparisons, and materialization trade-offs. More broadly, our results suggest that learned structures need not remain opaque objects invoked from SQL; they can become persistent, queryable components of the database itself. Doing so opens new questions in maintenance, physical design, optimization, and model lifecycle support. SOMs provide a concrete first case, but the broader opportunity is database-native learned analytical structures, including vector quantization, graph embeddings, and manifold summaries.



\bibliographystyle{ACM-Reference-Format}
\bibliography{references}

\end{document}